\documentclass[aps,prl,twocolumn,floatfix]{revtex4}

\usepackage{graphicx}
\usepackage{amsmath}

\begin{document}

\title{Entropy Production in High Energy Processes}
\author{Berndt M\"uller}
\affiliation{Department of Physics, Duke University, 
             Durham, North Carolina 27708} 
\author{Andreas Sch\"afer}
\affiliation{Institut f\"ur Theoretische Physik,
             Universit\"at Regensburg,
             D-93040 Regensburg, Germany}
\date{\today}

\begin{abstract}
We calculate the entropy produced in the decoherence of a classical
field configuration and compare it with the entropy of a fully
thermalized state with the same energy. We find that decoherence alone
accounts for a large fraction of the equilibrium entropy when each 
field mode is only moderately occupied. We apply this to theories
of relativistic heavy ion collisions, which describe the initial
state as a collection of coherent color fields. Our results suggest
that decoherence may partly explain the rapid formation of a high 
entropy state in these collisions.
\end{abstract}

\maketitle
Recent measurements of anisotropic flow in Au+Au collisions at the
Relativistic Heavy Ion Collider (RHIC) point to a very rapid formation 
of nearly equilibrated hot matter in these processes \cite{v2exp,v2th}.
Hydrodynamic calculations of the collective flow pattern yield estimates
for the equilibration time $\tau_{\rm eq} < 1$ fm/$c$. This is difficult
to understand in microscopic theories of thermalization \cite{biro,baier}, 
which suggest that the thermalization of the dense matter proceeds through 
radiative processes and takes several fm/$c$.

The initial state of a relativistic heavy ion collision is characterized
by a highly coherent configuration of quark and gluon fields. For the
processes contributing to the formation of matter near the center of
momentum of the two nuclei, coherent gluon fields at small values of
the Bjorken variable $x$ are most important. These fields are generated
by the quasistatic color charges of the valence quarks of the nuclei
and can be approximated as randomly oriented, quasiclassical color fields, 
often called a color glass condensate \cite{CGC}. During the nuclear
collision gluons are scattered out of this coherent field with a 
probability that is predicted to be close to unity \cite{Kov,Kras}. 
After their liberation, the gluons scatter off each other and radiate
additional gluons until they reach an equilibrium \cite{baier}.

Parton cascade models \cite{PCM} describe the liberation and rescattering
of gluons and other partons in a probabilistic framework based on the
relativistic Boltzmann equation, starting from an incoherent ensemble
of partons. On the other hand, models based on an initial state of the 
color glass condensate type treat the equilibration process as nonlinear 
classical evolution of the initial random, but coherent color fields 
\cite{Kras,Kovner,Poschl}. 
While entropy production in statistical transport theories, such as
the Boltzmann equation, is a well studied and understood phenomenon, 
the production of entropy by the decoherence of classical fields is 
less well understood. One mechanism for the production of entropy
is the pair creation of particles in strong (chromo-)electric fields 
\cite{Schwinger}. Another mechanism is the dynamical chaos generated
by the nonlinear field equations \cite{Chaos}, where the Kolmogorov-Sinai
(KS) entropy describes the apparent rate of entropy production.

Here we are not concerned with the microscopic description of the
production of entropy; instead, we address the question of the relative
contribution to entropy production by (a) the decoherence of an initially 
coherent field configuration and (b) the rescattering among incoherent
particle-like field excitations, which ultimately leads to equilibrium.
Since neither the initial color field configuration in a fast moving
nucleus nor its dynamical evolution after the collision of two nuclei 
is very well known, we start with a simple case, for which the relevant
calculations can be performed exactly, but which is sufficiently
general to permit conclusions that can be applied to heavy ion reactions.

The idea that decoherence may play a major part in entropy production 
in heavy ion collisions is not new. The notion of a large nucleus acting as
a ``phase filter'', decomposing the quark-gluon wavefunction of a hadronic
projectile into its incoherent components, was suggested over a decade ago 
\cite{LM91,Divonne}. The principles of decoherence of hadronic wavefunctions
were investigated extensively by Elze \cite{Elze} in the mid-1990s. The 
formulation of models for the small Bjorken-$x$ components of hadronic 
wavefunctions as superpositions of classical color fields, e.\ g.\ the
color glass condensate (CGC) model \cite{CGC}, now provides the theoretical 
basis for a more concrete treatment. In this framework we need to explore 
the effects of the decoherence of quasiclassical, coherent color fields,
which are present in the nuclei before the onset of a collision.

The quantum mechanical analogue of a classicical field is a coherent
state \cite{Glauber}:
\begin{equation}
| \Psi[J] \rangle = \prod_{\bf k} \exp( i\alpha_{{\bf k}\lambda} 
          a_{{\bf k}\lambda}^\dagger  - i\alpha_{{\bf k}\lambda}^* 
          a_{{\bf k}\lambda} ) | 0 \rangle ,
\label{eq1}
\end{equation}
where the amplitude $\alpha_{{\bf k}\lambda}$ is determined by the classical
current ${\bf J}$ creating the field:
\begin{equation}
\alpha_{{\bf k}\lambda} = (\hbar\omega_{\bf k}V)^{-1/2} 
          {\bf \epsilon}_{{\bf k}\lambda} \cdot 
          {\bf J}({\bf k},\omega_{\bf k}) .
\label{eq2}
\end{equation}
Let us begin by considering a single mode ${\bf k}\lambda$. The coherent
state can be written as a superposition of particle number eigenstates:
\begin{equation}
|\alpha\rangle = e^{-|\alpha|^2/2} \sum_{n=0}^\infty 
          \frac{\alpha^n}{\sqrt{n!}} |n\rangle .
\label{eq3}
\end{equation}
Being a pure quantum state, $|\alpha\rangle$ is described by a density
matrix
\begin{equation}
\rho_{mn} = \langle m|\alpha\rangle \langle\alpha |n\rangle ,
\label{eq4}
\end{equation}
which satisfies the relation $\rho^2 = \rho$ and has no entropy:
$S = - {\rm Tr}\,\rho\, \ln\rho = 0$. 

Complete decoherence of this quantum state corresponds to the total 
decay of all off-diagonal matrix elements of the density matrix, 
yielding the diagonal density matrix
\begin{equation}
\rho^{\rm dec}_{mn} = \langle n|\alpha\rangle|^2 \delta_{mn} =
          e^{-|\alpha|^2} \frac{|\alpha|^2n}{n!} \delta_{mn} .
\label{eq5}
\end{equation}
The particle number in this mixed state follows the Poisson distribution,
and the average number of particles is ${\bar n} = |\alpha|^2$. 
The entropy content of the mixed state is given by
\begin{eqnarray}
S_{\rm dec}^{\rm (cs)} & = & \sum_{n=0}^{\infty} 
          e^{-\bar n}\frac{{\bar n}^n}{n!}
          \ln\left(e^{-\bar n}\frac{{\bar n}^n}{n!}\right)
          \nonumber \\
          & = & e^{-{\bar n}} \sum_{n=0}^\infty \frac{{\bar n}^n}{n!} 
          (n \ln {\bar n} - {\bar n} - \ln n!) ,
\label{eq6}
\end{eqnarray}
where the superscript ``cs'' indicates that the result holds for a
coherent state. With the help of Stirling's formula and the integral 
representation of the logarithm,
\begin{equation}
\ln n = \int_0^{\infty} \frac{ds}{s} \left( e^{-s} - e^{-ns} \right) ,
\label{eq7}
\end{equation}
the sum in (\ref{eq6}) can be performed yielding an analytical result 
that is valid asymptotically for ${\bar n} \gg 1$ (actually, the
approximation is excellent already for ${\bar n} \approx 1$):
\begin{equation}
S_{\rm dec}^{\rm (cs)} = \frac{1}{2} \left( \ln (2\pi {\bar n}) \, + 1 -
          \frac{1}{6{\bar n}} + \cdots \right) .
\label{eq8}
\end{equation}
It is not surprising that the entropy is proportional to 
$\ln \sqrt{\bar n}$, because we have deleted all information about 
the relative signs of the amplitudes $\langle\alpha |n\rangle$ by 
eliminating the off-diagonal elements of the density matrix. 
The number of significantly contributing elements is given by the
width, $\Delta n = \sqrt{\bar n}$, of the Poisson distribution.

\begin{figure}[tb]   
\resizebox{0.95\linewidth}{!}
          {\rotatebox{-90}{\includegraphics{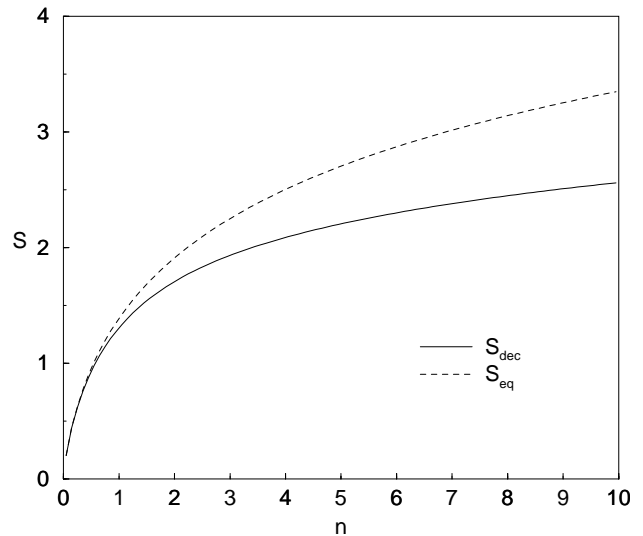}}}
\caption{Decoherence entropy $S_{\rm dec}$ for a coherent state of a 
         single field mode and equilibrium entropy $S_{\rm eq}$ for 
         the same average total energy as a function of the average 
         occupation number $\bar n$.}
\label{fig1}
\end{figure}

Let us mention that the energy for a single quantum oscillator in
equilibrium at temperature $T$ is given by
\begin{equation}
S_{\rm eq} = \ln ({\bar n}+1) + 
             {\bar n}\ln\left(1+\frac{1}{\bar n}\right) ,
\label{eq9}
\end{equation}
where ${\bar n} = (e^{\omega/T}-1)^{-1}$ is the average occupation
number. Asymptotically, for large $\bar n$, one obtains $S_{\rm eq}
\approx 2 S_{\rm dec}^{\rm (cs)}$, i.\ e.\ the thermal entropy becomes
twice as large as the decoherence entropy. However, for small to
moderate occupation numbers the ratio $S_{\rm dec}^{\rm (cs)}/S_{\rm eq}$ 
is close to unity and remains above 0.75 up to ${\bar n} = 10$. 
Figure \ref{fig1} shows the decoherence and equilibrium entropies 
as a function of the average occupation number $\bar n$. It is 
evident that, for not too large values of $\bar n$, the decoherence 
process generates a large fraction of the entropy that can be created,
and any subsequent equilibration process adds only a small amount of
entropy to it. Since decoherence is usually a much faster process
than thermal equilibration, our result implies that the fast entropy
production observed in heavy ion collisions may be primarily due to
decoherence of the initial state color fields.

What does this imply for the quantum field theory, where the field is
a system of (infinitely) many coupled oscillators? Assume that, after 
decoherence, the system can be described as a collection of $N$
particles, given by some distribution function over single-particle
states, which were generated by the decoherence of $N_{\rm cs}$ 
coherent quantum states. Examples of such states are the internal
wavefunctions of nucleons forming a large nucleus, or a quark with
its comoving gluon cloud. Each coherent state contributes on average
${\bar n} = N/N_{\rm cs}$ partons. Then, after full equilibration,
the thermal entropy is of the order of $S_{\rm th} \sim N_{\rm cs}
{\bar n} = N$, while for the decoherence entropy we get
$S_{\rm dec} \sim N_{\rm cs} \frac{1}{2} \ln(2\pi{\bar n})$.
The ratio of the two entropies is
\begin{equation}
\frac{S_{\rm dec}}{S_{\rm th}} \sim \frac{\ln(2\pi{\bar n})}{2 \bar n} ,
\label{eq10}
\end{equation}
i.\ e.\ for large amplitude quantum states, which turn into many
particles per coherent mode, the decoherence contribution to the
thermal entropy is small. On the other hand, if the individual
occupation numbers are of order one, the contribution is sizable.
This case applies to our problem of interest, the collision of two 
nuclei at high energy, as we will discuss below.

For the coherent color fields in colliding nuclei, the average number
of decohering gluons per transverse area has been given by \cite{Kov}
\begin{equation}
\frac{dN}{d^2b} \approx
           \frac{C_{\rm F}\ln 2\, Q_s^2\Delta y}{\pi^2\alpha_s} ,
\label{eq11}
\end{equation}
where $Q_s$ is the so-called saturation scale, $C_{\rm F}=4/3$ is the 
quadratic Casimir operator in the fundamental representation of SU(3), 
and $\Delta y$ is the rapidity interval over which the color fields 
retain their coherence. The characteristic transverse area, over 
which the color fields are coherent, is $\pi/Q_s^2$, and one can
argue that the longitudinal coherence length is of the order of
$\Delta y \approx 1/\alpha_s$ \cite{KLM01}. We thus obtain an average
number of decohering partons per coherence region:
\begin{equation}
{\bar n} \approx \frac{C_{\rm F}\ln 2}{\pi\alpha_s^2} \approx 3 .
\label{eq12}
\end{equation}
For this value, our arguments presented above indicate that the entropy 
produced in the decoherence process is about half of the equilibrium
entropy. The total entropy per unit rapidity produced by decoherence 
in a Au+Au collision at the Relativistic Heavy Ion Colider is 
\begin{eqnarray}
\frac{dS_{\rm dec}}{dy} & \approx & \frac{Q_s^2 R^2}{2 \Delta y} 
          \left(\ln(2\pi {\bar n}) +1\right)
          \nonumber \\
          & \approx & \frac{1}{2} Q_s^2 R^2 \alpha_s \left[
          \ln \frac{2C_{\rm F}\ln 2}{\alpha_s^2} + 1 \right]
          \approx 1500 ,
\label{eq13}
\end{eqnarray}
where we used the values \cite{Kov} $Q_s^2\approx 2$ GeV$^2$, $R=7$ fm,
and $\alpha_s \approx 0.3$. This value accounts for about half of the
entropy measured in the final hadron distribution.

Entropy production by decoherence of classical color fields was also 
discussed recently by A.~Mueller on the basis of somewhat different
arguments \cite{Al}, who obtained a similar expression for the total 
generated entropy:
\begin{equation}
S_{\rm dec} \sim c_S Q_s^2R^2 ,
\end{equation}
where $c_S$ is a nontrivial factor numerical of order one. It is the 
factor $Q_s^2R^2$, which makes entropy generation by decoherence 
a large effect in his treatment, as well. Our results for the numerical 
factor differ in their dependence on $\alpha_s$, but not in their order 
of magnitude. An unambiguous determination of the factor $c_S$ will 
require a determination of the (de-)coherence length for the classical 
fields, which exist in the nuclei before the collision.

We finally discuss the case of particle production by a fast moving 
electric charge, such as a large nucleus. The occupation number of the
various field modes forms the basis of the Weizs\"acker-Williams (WW)
approximation to interactions of charged particles and photons at
high energies \cite{WW}. For a Coulomb charge $Ze$ moving close to the
speed of light with Lorentz factor $\gamma$, the spectrum of equivalent
photons is given by 
\begin{equation}
\frac{dn}{d\omega} = \frac{2Z^2\alpha}{\pi\omega} 
                     \ln \left(\frac{\gamma}{\omega R}\right) ,
\label{eq14}
\end{equation}
where $R$ is the intrinsic size of the charge and 
$\omega < \omega_{\rm max} = \gamma/R$.
In order to determine the occupation number for a photon energy $\omega$,
we need to know the coherence interval $\Delta\omega$, corresponding to
the inverse longitudinal length scale of the process which causes the
destruction of the coherent field. Characteristically, $\Delta\omega$
grows with $\omega$. For our analysis we assume that $\Delta\omega
\sim \epsilon\omega$ with a constant parameter $\epsilon << 1$ and
discretize the integral over $\omega$ by setting $\omega_j = 
\omega_{\rm min}e^{\epsilon j}$ with integer $j=1,\ldots,J$ and
$J=\epsilon^{-1}\ln(\omega_{\rm min}/\omega_{\rm max})$.  
The entropy generated by decoherence of the initial field configuration 
is then obtained as
\begin{eqnarray}
S_{\rm dec} & \approx & \sum_{j=1}^J \frac{1}{2}\ln\left[ 4Z^2\alpha\epsilon 
            \ln\left( e^{-\epsilon j}\frac{\omega_{\rm max}}{\omega_{\rm min}}
            \right) \right]
            \nonumber \\
            & \approx & \sum_{j=1}^J \frac{1}{2}\ln\left[ 4Z^2\alpha\epsilon^2 
            (J-j) \right]
            \nonumber \\
            & \approx & \frac{1}{2\epsilon} 
            \ln\frac{\omega_{\rm max}}{\omega_{\rm min}}
            \cdot \ln\left( 4Z^2\alpha e^{-1}\epsilon 
            \ln\frac{\omega_{\rm max}}{\omega_{\rm min}} \right) .
\label{eq15}
\end{eqnarray}
Thus the produced entropy depends crucially on the value of $\epsilon$,
and thus on the process leading to decoherence. In the case of coherent
color fields, a more quantitative understanding of the coherence length
of the fields before and after the collision is desirable.

In conclusion, we have shown that the decoherence of a quasiclassical state
generates a significant amount of entropy. If the average occupation number
of each coherent domain of the initial state is not much larger than one,
the entropy released by the decoherence process is a sizable fraction 
(e.\ g.\ one-half) of the entropy attained after thermodynamic equilibration.
If this reasoning is applied to relativistic heavy ion collisions in the
framework of the color glass condensate model, a significant fraction of
the measured entropy of the final state can be produced on the time scale
of decoherence, $\tau_{\rm dec} \sim 1/Q_s < 0.2$ fm/$c$. This may
explain the rapid transition to a state that behaves approximately like
an equilibrated QCD plasma. The precise value of the entropy generated
by decoherence (\ref{eq13}) depends sensitively on the coherence length
of the entropy creating process.

\begin{acknowledgments}  
We thank H.~T.~Elze for helpful comments on an earlier version of
this manuscript.
This work was supported in part by grants from the U.S. Department
of Energy and the BMBF, and by the Alexander von Humboldt Foundation.
\end{acknowledgments}

\end{document}